\documentclass[reprint,superscriptaddress,amsmath,amssymb,aps]{revtex4-1}
\usepackage[utf8]{inputenc}
\usepackage[T1]{fontenc} 
\usepackage{graphicx}
\usepackage{rotating}
\usepackage{makeidx}
\usepackage{threeparttable}
\usepackage{float}
\usepackage{amsmath}
\usepackage{rotating}
\usepackage{hyperref}

\usepackage{relsize}
\usepackage{etoolbox}
\usepackage[most]{tcolorbox}
\usepackage{hyperref}
\usepackage{dcolumn}
\usepackage{bm}
\usepackage{wrapfig}
\usepackage{float}  
\usepackage[toc,page]{appendix}
\usepackage{gensymb}
\usepackage{epstopdf}
\usepackage{subfiles}
\usepackage{comment} 
\usepackage[autostyle=true]{csquotes} 
\usepackage{makecell}
\usepackage{siunitx}
\usepackage[all]{hypcap} 
\usepackage[most]{tcolorbox}
\usepackage{hyperref}
\usepackage{xcolor}
\usepackage{nicefrac}
\usepackage{chemformula}
\usepackage{ulem}  
\definecolor{gainsboro}{rgb}{0.86, 0.86, 0.86}
\usepackage[backgroundcolor=gainsboro]{todonotes}
\hypersetup{
  colorlinks   = true, 
  urlcolor     = blue, 
  linkcolor    = blue, 
  citecolor   = blue 
}

\newcommand{\weg}[1]{}

\newcommand{\obyth}[2]{$\frac{1}{3}$}
\newcommand{\obyf}[2]{$\frac{1}{4}$}

\begin{document}

\title{Ultrafast spin-nematic and ferroelectric phase transitions 
induced by femto-second light pulses}

\author{Sangeeta Rajpurohit}
\email{srajpurohit@lbl.gov}
\affiliation{Molecular Foundry, Lawrence Berkeley National Laboratory, USA}

\author{Liang Z. Tan}
\affiliation{Molecular Foundry, Lawrence Berkeley National Laboratory, USA}

\author{Christian Jooss}
\affiliation{Institute for Material Physics, Georg-August-Universit{\"a}t G{\"o}ttingen, Germany}

\author{P. E. Bl{\"o}chl}
\affiliation{Institute for Theoretical physics, Clausthal University of Technology, Germany}
\affiliation{Institute for Theoretical Physics, Georg-August-Universit{\"a}t G{\"o}ttingen, Germany}


\begin{abstract}
Optically-induced phase transitions of the manganite $\rm Pr_{1/3}Ca_{2/3}MnO_3$ have been simulated using a
model Hamiltonian, that captures the dynamics of strongly correlated charge, orbital, lattice and spin degrees of freedom. 
Its parameters have been extracted from first-principles calculations. Beyond a critical intensity of a femto-second light pulse,
the material undergoes ultra-fast and non-thermal magnetic phase transition from a non-collinear to collinear antiferromagnetic phases.
The light-pulse excites selectively either a spin-nematic or a ferroelectric phase depending on the light-polarization.  The behavior
can be traced to an optically induced ferromagnetic coupling between Mn-trimers, i.e. polarons which are delocalized over three Mn-sites.  
The polarization guides the polymerization of the polaronic crystal into distinct patterns of ferromagnetic chains determining the target phase.
\end{abstract}
\maketitle

\section{Introduction}
The manipulation of local spin order in a magnetic material by electrical or optical means forms the basis of proposed "beyond Moore's Law"
information technologies. Photo-induced magnetic phase transitions are possibly the fastest way to alter the spins in magnetic materials, as
demonstrated by femtosecond timescale experiments in several ferromagnetic (FM) systems \cite{Beaurepaire1996, Hohlfeld1997,Gudde1999,Kirilyuk2010}.
While direct access to the magnetic order in antiferromagnetic and non-collinear spin systems by experiments has been challenging due
to very weak or zero net magnetic moments, recent advances have opened new possibilities to use these spin orders as information systems:
Optically-induced magnetic phase-transitions have been observed from antiferromagnetic to paramagnetic states in $\rm FeBO_3$ \cite{Kimel2002},
from collinear to non-collinear antiferromagnetic states in $\rm {DyFeO_3}$ \cite{Afanasiev2016} and CuO\cite{Johnson2012}.  
Inducing weak ferromagnetism through Dzyaloshinskii–Moriya interactions by optically altering local spins in antiferromagnets can
be crucial in multiferroicity.   
 
Perovskite manganites are a class of materials with strongly correlated spin, lattice, charge and orbital degrees of freedom.\cite{Jirak1985,Schiffer1995,Tokura2006_2,Jooss2007}
This results in a rich phase diagram with long-range ordering patterns. Recently, several fascinating photo-induced phenomena such
as magnetic phase transitions, hidden phases, and long-lived excitations have been observed experimentally in the charge- and
orbital-ordered manganites \cite{Li2013,Beaud2014,Raiser2017,Lin2018}. The experimental studies of optical manipulation
of spin orders in manganites show these systems are promising materials for opto-spintronics \cite{Fiebig2008,Li2013,Beaud2014,Satoh2015,Manz2016}. 
The understanding of the physical mechanisms governing the ultrafast spin dynamics is crucial in view of later applications.

The intrinsic mechanisms that govern the photo-induced magnetic phase transitions are not yet fully understood in manganites.
The theoretical understanding of the photo-induced phase transitions requires microscopic knowledge of the energy conversion processes.
The charge, spin, and lattice degrees are known to actively participate in the energy conversion processes in manganites \cite{Tokura2006_2,Jooss2007,Li2013,Raiser2017}.
Studying the photo-excitation and its subsequent relaxation in manganites through theoretical models is challenging. The time and length scales
of the phonon and spin dynamics in manganites are beyond the scope of the traditional first-principles approaches, such as time-dependent
density functional theory (TD-DFT) \cite{Runge1984}, usually employed to study the nonlinear optical response in solids.

In the present paper, we report on simulations of the photo-excitation and the subsequent non-equilibrium relaxation
dynamics of the  perovskite manganite ${\rm Pr_{1/3}Ca_{2/3}MnO_3}$. We find optically induced, non-thermal phase transitions
from a non-collinear spin-order into a spin-nematic or a ferroelectric phase.

Simulations have been performed  on a pico-second time scale using a tight-binding model that captures the correlated
dynamics of electronic, spin and lattice degrees of freedom with Ehrenfest dynamics.

In the ground state, $\rm Pr_{1/3}Ca_{2/3}MnO_3$ has a charge- and orbital-ordered stripe-phase with non-collinear spin order.
The magnetic phase transition is observed above a critical intensity of the light pulse. Two distinct sets of antiferromagnetic patterns
can selectively be produced by choosing the polarization direction of the light pulse:  One set of optically induced phases exhibits ferroelectricity. 
The other set are novel spin-nematic phases with two-fold rotational symmetry. It will be discussed how to selectively drive the system into
specific broken-symmetry states and ways to probe them will be suggested. The detailed analysis provides a fairly general picture
 of optically induced magnetic phase transitions.

\begin{figure}[t]
\begin{center}
\includegraphics[width=\linewidth]{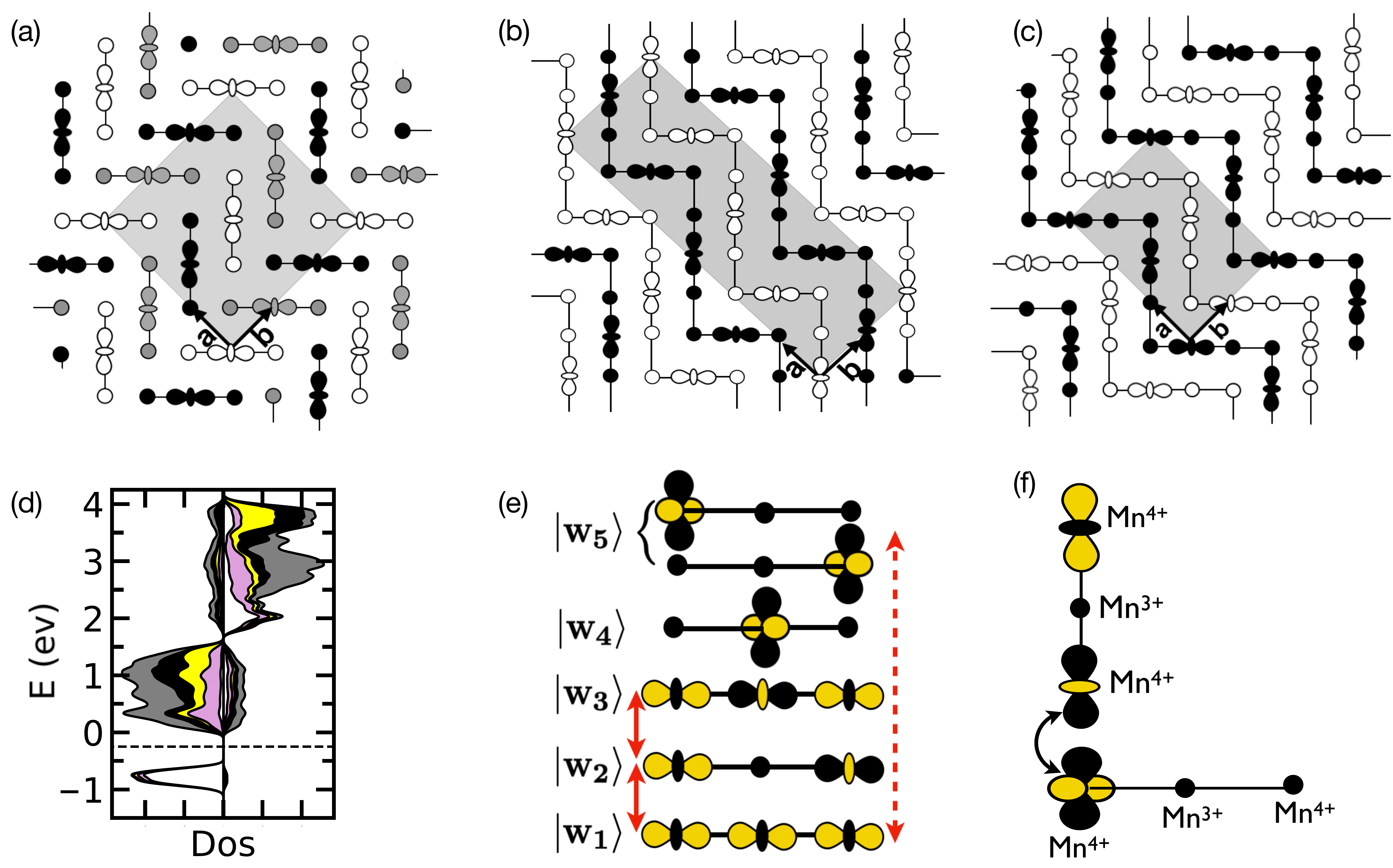}
\end{center}
\caption{\label{fig1}Charge, orbital, and spin order 
of $\rm Pr_{1/3}Ca_{2/3}MnO_3$ in (a) the ground state, (b) the spin-nematic phase obtained with a polarization along
$\vec{a}+\vec{b}$  and (c) the ferroelectric phase obtained with a polarization along $\vec{a}$ or $\vec{b}$. The
three coplanar spin directions at Mn sites are shown in white, gray, and black.  (d)\label{fig:dos} Density of states projected
on the Wannier states $|w_1\rangle$ (white), $|w_2\rangle$ (pink), $|w_3\rangle$ (yellow), $|w_4\rangle$ (black) and $|w_5\rangle$ (grey). 
The right- and left-side shows contribution from majority- and minority-spin states and the dashed line is the Fermi level. 
(e)\label{fig:wannierstates} Wannier states $|w_i\rangle$ with $i\in\{1,2,...5\}$ localized on a Mn-trimer. 
The red solid arrows indicate dipole-allowed intra-trimer electronic transitions.
(f)\label{fig:w2w5} $|w_2\rangle$-$|w_5\rangle$ state hopping of adjacent orthogonal trimers, shown by red dashed arrow in (e).}
\end{figure}
\section{Theoretical details}
\subsection{Model Hamiltonian}
To investigate the charge, spin and orbital orders in perovskite manganites, we use a tight-binding model, which has been
 described in detail in refs.~\cite{Sotoudeh2017,Rajpurohit2020}. Here we provide a brief summary of the model.

In $\rm Pr_{1-x} Ca_{x}MnO_3$, the octahedral crystal-field splits the Mn 3d-shell into three non-bonding $\rm t_{2g}$
 orbitals and two antibonding $e_g$ states, which are shifted up in energy.

The electron subsystem of the model consists of $\rm e_g$-electrons. The $\rm e_g$-electrons are described
 by the one-particle wave functions
\begin{eqnarray}
|\psi_n\rangle=\sum_{\sigma,\alpha,R}
|\chi_{\sigma,\alpha,R}\rangle
\psi_{\sigma,\alpha,R,n}
\label{eq:psin}
\end{eqnarray}
with band index $n$. 
The wave functions are expressed in the basis set of local Mn eg-orbitals $|\chi_{\sigma,\alpha,\vec{R}}\rangle$ with
orbital index $\alpha$ ($\alpha\in\{x^2-y^2,3z^2-r^2\}$), spin index $\sigma$ ($\sigma\in\{\uparrow,\downarrow\}$) and Mn-site index $R$.
The complex-valued orbital expansion coefficients are $\psi_{\sigma,\alpha,R,n}$.

The corresponding local one-particle reduced density matrix ${\rho}_R$ at site $R$ is defined by
\begin{eqnarray}
\rho_{\sigma,\alpha,\sigma',\alpha',R}{=}\sum\limits_{n}
 f_n\psi_{\sigma,\alpha,R,n}\psi^*_{\sigma',\alpha',R,n}.
 \label{eq:dens_mat}
 \end{eqnarray}
where $f_n$ are the occupations.
 
The $t_{2g}$ electrons in manganites are fully spin polarized and remain localized \cite{Sotoudeh2017}.
 The $t_{2g}$ electrons are represented by a spin vector $\vec{S}_R$ with length $\frac{3}{2}\hbar$ on each Mn site.
While the spin of the ${\rm t_{2g}}$ electrons are described by a spin $\vec{S}_R$, the spin of the $\rm{e_g}$-electrons
is described directly by the wave functions, which are Pauli spinors.

The phonon subsystem of an oxygen octahedron centered at site $R$ consists of three vibrational modes,
namely two Jahn-Teller active modes, $Q_{2,R}$ and $Q_{3,R}$, as well as a breathing mode $Q_{1,R}$. These phonon
modes are expressed in terms of the displacements of the oxygen ions along the oxygen bridge. Because the oxygen
ions are, each, shared between two MnO$_6$ octahedra, the phonon modes are highly cooperative. Our model takes this effect fully into account.

The potential energy functional of our tight-binding model of a manganite is
\begin{eqnarray}
E_{pot}\Big(|\psi_n\rangle,\vec{S}_{R},Q_{i,R}\Big)
&=&E_{e}(|\psi_n\rangle) +E_{S}(\vec{S}_R)+E_{ph}(Q_{i,R}) \nonumber\\
&&\hspace{-2cm}+
E_{e-ph}(|\psi_n\rangle,Q_{i,R}+E_{e-S}(|\psi_n\rangle,\vec{S}_R)
 \label{eq:tbm}
\end{eqnarray} 
expressed in terms of electronic degrees of freedom $|\psi_n\rangle$, spins $\vec{S}_R$ and lattice degrees of freedom $Q_{i,R}$.

The energies $E_{e}$, $E_{S}$ and $E_{ph}$ in Eq.~\ref{eq:tbm} are those of the separated $\rm{e_g}$-electron, spin, and phonon subsystems, respectively. 
$E_{e-ph}$ is the electron-phonon coupling and $E_{e-S}$ is the Hunds-rule coupling between the spins of ${\rm e_g}$ and ${\rm t_{2g}}$ electrons on the same site. 
The energy $E_e$ of electron subsystem  is due to the kinetic energy of the $e_g$ electrons and the onsite Coulomb-interaction between $e_g$ electrons.
The energy $E_S$ of the spin sub-system describes  the Heisenberg-type antiferromagnetic coupling between the spins on neighboring Mn-sites.
The energy $E_{ph}$ of the phonon-subsystem includes the restoring energy, which is quadratic in the mode amplitudes.

As the perovskite systems are described in the $Pbnm$ space group, we use the corresponding lattice vectors with
 $\vec{a}||(\vec{e}_y{-}\vec{e}_x)$, $\vec{b}||(\vec{e}_x{+}\vec{e}_y)$ and $\vec{c}||\vec{e}_z$, where $\hat{e}_x$, $\hat{e}_y$ and $\hat{e}_z$
 are the directions pointing along the nearest Mn-Mn sites.
  
\subsection{Dynamics}
To study the dynamics of the system, we combine the above model in Equation \ref{eq:tbm} with Ehrenfest dynamics. 
In the spirit of Ehrenfest dynamics, we propagate the single-particle wave functions $|\psi_{\sigma,\alpha,R,n}\rangle$ and the
spins $\vec{S}_R$ with the time-dependent Schrodinger equation (TDSE), while the atoms are treated as classical particles
that follow Newton's equations of motion.  The spin vectors $\vec{S}_R$, which describe $t_{2g}$ states, are expressed by
two-dimensional spinors, which enforces strict spin alignment of the $t_{2g}$-spins on a given site.  
This framework can be considered as implementation of time-dependent density functional theory in the adiabatic
approximation for the specified model Hamiltonian \cite{Runge1984}.

The optical excitation has been implemented using Peierls substitution, which maps the time dependent vector
potential of the light pulse onto a modulation of the hopping parameters of the electrons.

The details of our implementation of Ehrenfest dynamics and the optical excitaion have been given earlier\cite{Rajpurohit2020}.

\section{Results and discussion}
The ground state of $\rm Pr_{1/3} Ca_{2/3}MnO_3$ is the so-called stripe phase.\cite{Radaelli1999}.
Its charge, orbital and spin order is shown in figure~\ref{fig1}.a.
Although the charge and orbital order of the stripe-phase $\rm Pr_{1/3} Ca_{2/3}MnO_3$ is well established, its spin order
is yet under debate\cite{Jirak1985,Radaelli1999,Jirak2005}.  Based on our calculations, we propose the new spin order,
shown in figure~\ref{fig1}.a, which has a lower energy than those suggested previously \cite{Radaelli1999,Jirak2005,Hotta2000_4} 
\footnote{The previously proposed collinear \cite{Hotta2000_4} and non-collinear\cite{Jirak1985, Radaelli1999,Jirak2005} spin orders have been used as initial states for the optimization.}. 

The ground state has a coplanar spin order with three distinct spin axes with an angle of $120\degree$ among each other.
The spin order in the $ab-$planes can be understood as an arrangement of trimers. 
Each trimer is a segment of three ferromagnetically aligned Mn sites in a row along $\vec{a}+\vec{b}$ or
$\vec{a}-\vec{b}$ directions in the $ab$ plane.  Each trimer is surrounded in the $ab$-plane
by neighboring trimers with a relative spin angle of either $+120\degree$ or $-120\degree$.  
The $ab$-planes are antiferromagnetically coupled along the $c-$axis.    

The central Mn-ion of a trimer has a $\rm Mn^{3+}$ oxidation state and the terminal Mn-ions are in the
$\rm Mn^{4+}$ oxidation state. Oxidation states are integer by definition. 
The real charge distribution is more subtle: In our tight-binding model, the formal $\rm Mn^{3+}$-ions
have $0.71$~$e_g$ electrons, while the formal $\rm Mn^{4+}$-ions have $0.145$~$e_g$ electrons. 
  
The Jahn-Teller effect lifts the degeneracy of the $e_g$ orbitals at the central $\rm Mn^{3+}$ site and
simultaneously distorts its MnO$_6$ octahedron.  The orbital-polarization at $\rm Mn^{3+}$ sites is of $d_{3x^2-r^2}$, respectively $d_{3y^2-r^2}$-type,
\footnote{The occupied $e_g$ orbitals at Mn$^{3+}$ sites are linear combination of $e_g$ states
$|\theta\rangle{=}-|x^2-y^2\rangle \sin(\theta) +|3z^2-r^2\rangle\cos(\theta)$ with $\theta{=}60\degree$
which suggests almost $d_{3x^2-r^2}$/$d_{3y^2-r^2}$-type orbital-polarization}
and forms a long range orbital-order pattern shown in figure \ref{fig1}.a.  The Jahn-Teller distortion at the
formal $\rm Mn^{4+}$ ions is smaller, consistent with their $e_g$ occupancy. 
  
Analogously to the half-doped $\rm Pr_{1-x}Ca_{x}MnO_3$,\cite{Sotoudeh2017} the electronic structure
can be rationalized by Wannier states centered on the trimers. They are shown in Figure \ref{fig1}.e.

The filled states are well represented by the $|w_1\rangle$ Wannier states as seen in the projected density of states \ref{fig1}.d.
They have bonding character and a large weight on the Mn$^{3+}$ central sites.  Above the Fermi level are the non-bonding
$|w_2\rangle$-states which are localized on the terminal sites of the trimer.  The $|w_2\rangle$ states can strongly hybridize
with the nearly iso-energetic $|w_5\rangle$ orbitals of a neighboring trimer. The Wannier states with highest energy are the fully
antibonding trimer state $|w_3\rangle$ and the upper Jahn-Teller orbital $|w_4\rangle$ of the central site.

\subsection{Photo-induced magnetic phase transitions}
To investigate the photo-excitation, we choose a linearly polarized 100-fs light pulse with photon energy $\hbar\omega{=}0.92$~eV,
for which  the system exhibits maximum absorption.  The electromagnetic field associated with the pulse is
$\vec{E}(r,t){=}\vec{e}_A\omega\rm{Im}(A_oe^{-i\omega t})g(t)$, where $A_o$ is the amplitude of the vector potential,
$\omega$ is the angular frequency, $\vec{e}_{A}$ is the direction of the electric field and the Gaussian pulse shape
is imposed by $g(t){=}e^{-\frac{t^2}{2c_w^2}}(\sqrt[4]{\pi c^2_w})^{-1}$ which has FWHM of $2c_w\sqrt{\ln2}$. 
The effect of the electromagnetic field is incorporated in the model through the Peierls substitution as described
earlier\cite{Rajpurohit2020}.  A $12{\times}12{\times}4$ supercell with 576 Mn-ions has been used for the simulations.

To monitor the phase-transition, we use the spin-correlation function

\begin{eqnarray}
C_{S}(\vec{G}){=}\frac{1}{N} \biggl| \sum_{R=1}^N e^{i\vec{G}{\vec{R}_R}} (\vec{S}_R+\vec{s}_R)\biggr|^2
\end{eqnarray}
where $\vec{S}_{R_i}$ and $\vec{s}_{R_i}$ are the spins of
$t_{2g}$ and $e_g$ electrons, respectively, at site $R$.
The wave vector $\vec{G}$ is represented by its relative coordinates
$(h,k,l)$ in the $Pbnm$ setting. 

\begin{figure}[t]
\begin{center}
\includegraphics[width=\linewidth]{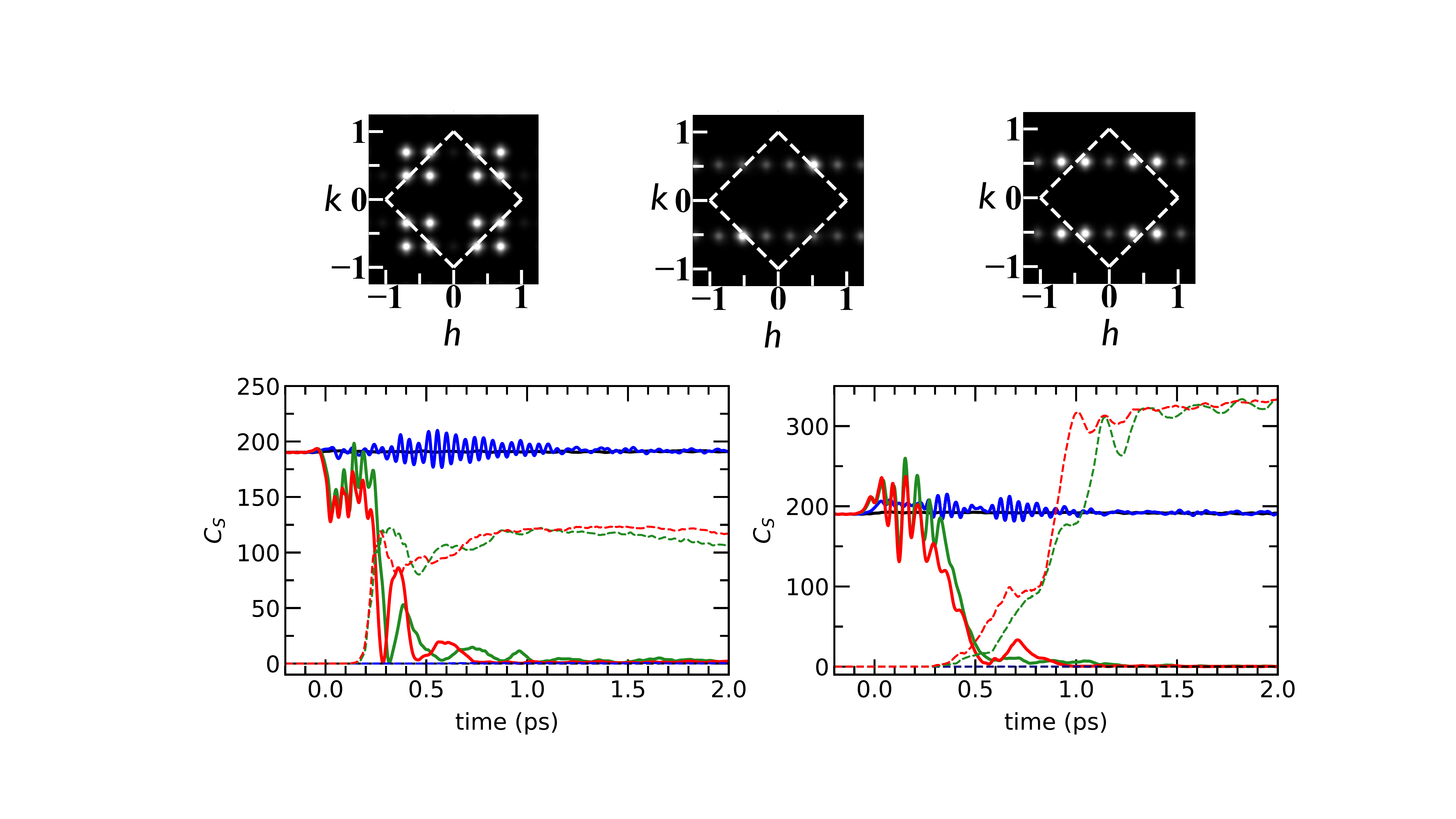}
\end{center}
\caption{\label{fig2}
Top: Spin correlation function $C_S$ for $l{=}1$ of the ground state (left), the nematic phase obtained by a light pulse
polarized along $\vec{a}\pm\vec{b}$ (middle) and the ferroelectric phase obtained with polarization along $\vec{a}$ or $\vec{b}$.
The correlation functions of the optically induced phases are shown for 1~ps after the pulse. Bottom: Time evolution of
the spin correlation function at $(1/2,1/2,1)$ characteristic for the spin-nematic phase (left). The contribution of the
diffraction peaks from the immediate neighborhood of $(1/2,1/2,1)$ peak is taken into account. Evolution of the spin
correlation function at (1/3,1/2,1), which is characteristic for the ferroelectric phase (right). The full lines show the spin
correlation at (1/3,1/3,1) characteristic for the ground state.
The colors refer to the increasing intensities with 
$A_o{=}$ $0.10$ $\hbar/ea_o$ (black), 
$0.225$ $\hbar/ea_o$ (blue),
$0.45$ $\hbar/ea_o$ (green) and 
$0.50$ $\hbar/ea_o$ (red).
The corresponding photon absorptions per
Mn $D_p$ for polarization along $\vec{b}$ $(\vec{a}+\vec{b})$ are 
$D_p$=0.002(0.001)~ph/Mn (black), 
0.016(0.015)~ph/Mn (blue), 
0.047(0.039)~ph/Mn (green) and  
0.062 (0.043) ph/Mn (red). } 
\end{figure}
The charge and orbital order is monitored by their correlation functions.
The charge-correlation $C_{Q}$ is  $C_{Q}(\vec{G}){=}\frac{1}{N} \big| \sum_{R=1}^N e^{i\vec{G}{\vec{R}_R}} (n_R-\langle n \rangle)\big|^2$, where $n_R{=}\sum_{\alpha,\sigma}\rho_{\sigma,\alpha,\sigma,\alpha,R}$ 
is the $e_{g}$-electron-density at site $R$, and $\langle n \rangle{=}1/3$ is its average value.

The orbital-correlation function $C_O(\vec{G})$ is $C_{O}(\vec{G}){=}\frac{1}{N} \big| \sum_{R=1}^N e^{i\vec{G}{\vec{R}_R}}  (n_{x,R}{-}n_{y,R})\big|^2$,
where $n_{x,R}$ and $n_{y,R}$ are the occupancies of two orthonormal $e_g$ states $|\theta_j\rangle$ with $j\in\{x,y\}$ at site $R$. 
The orbitals are $\theta_{x}=(|3z^2-r^2\rangle-|x^2-y^2\rangle)/\sqrt{2}$
and $\theta_{y}=(|3z^2-r^2\rangle+|x^2-y^2\rangle)/\sqrt{2}$, which are orthonormal and point predominantly in $x$
and $y$ direction, respectively.  They are chosen consistent with resonating x-ray diffraction experiments at the Mn-K
edge \cite{Zimmermann2001}, which involve 1s-to-4p states Mn transitions.

The occupancies $n_{j,R}$ of orbital $|\theta_{j}\rangle$ at site $R$ is given by
\begin{eqnarray}
n_{j,R}{=}\sum_{\sigma}\sum_n \langle\theta_{\sigma,j,R}|\psi_n\rangle f_n\langle\psi_n|\theta_{\sigma,j,R}\rangle
\end{eqnarray}
with the one-particle wave functions $|\psi_n\rangle$ from Eq.~\ref{eq:psin}
and orbitals $|\theta_{\sigma,j,R}\rangle$ centered at site $R$ with spatial $|\theta_{j}\rangle$
character and spin $\sigma$.

The dynamics of the system has been studied for four different polarization directions  of the light-pulse
in the $ab$-plane, namely with the electric field along $\vec{a}$, $\vec{b}$, $\vec{a}+\vec{b}$, and $\vec{a}-\vec{b}$. 

The excitation leads to two distinct outcomes depending on the polarization of the light pulse:
\begin{itemize}
\item A polarization along the oxygen bridges, i.e. parallel to $\vec{a}+\vec{b}$ or  $\vec{a}-\vec{b}$,
drives the material into a spin-nematic phase shown in figure~\ref{fig1}.b.  
\item A polarization along  $\vec{a}$ or $\vec{b}$ leads to a ferroelectric phase shown in figure~\ref{fig1}.c.
\end{itemize}

The spin-correlation functions for the ground state and the two optically-induced phases are shown in figure~\ref{fig2}. 
The order parameter for the transition is the spin-correlation function at specific reciprocal lattice vectors. 
Their time dependence demonstrating the ultra-fast response on a sub-picosecond time scale is shown in figure~\ref{fig2}.

\begin{itemize}
\item The coplanar spin order of the ground state displays strong peaks in the spin-correlation $C_S$ at
 wave vectors $(h,k,l)=(u\pm1/3,v\pm1/3,2w-1)$ with integer $u,v,w$ and independent $\pm$. 

\item The spin-nematic phases are characterized by two sets of $C_S$ peaks. The first set appears at wave
vectors $(h,k,l)=(u+1/2,v+1/2,2w-1)$ with integer $u,v,w$ with even $u+v$. On longer timescales, each diffraction
spot of the first set separates into two twin peaks located at the supercell reciprocal-space vectors adjacent 
to the first set of peaks. The second set, 4-5 times weaker, has wave vectors $(h,k,l)=(u\pm1/6,v-1/2,2w-1)$ with integer $u,v,w$.

\item The ferroelectric phases have strong
$C_S$ peaks at wave vectors $(h,k,l)=(u\pm1/3,v+1/2,2w-1)$ with integer $u,v,w$.
\end{itemize} 

Let us first describe the excitation by light polarized parallel to the oxygen bridges, that is along
 $\vec{a}+\vec{b}$ or  $\vec{a}-\vec{b}$. These polarizations  results in the spin-nematic phases.
The excitation takes place at those trimers that are aligned with the electric field. The excitation
lift electrons from the Wannier state $|w_1\rangle$, shown in figure~\ref{fig1}.e, to the antisymmetric state 
$|w_2\rangle$.  The corresponding dipole oscillation shuffles charge between the two terminal
Mn-sites of the excited trimer.

\begin{figure}[t]
\begin{center}
\includegraphics[width=\linewidth]{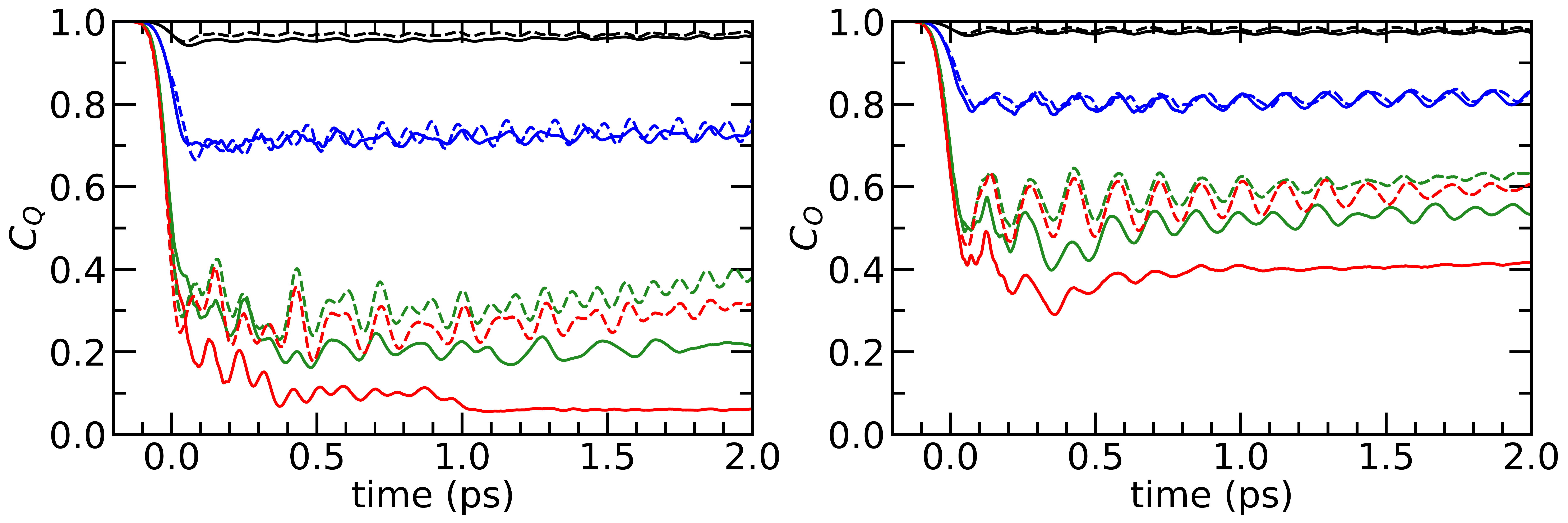}
\end{center}
\caption{\label{fig3}
Charge correlation $C_Q$ at $(h,k,l){=}(1/3,1,0)$ and orbital correlation $C_O$ at $(h,k,l){=}(1/3,0,0)$ as function
of time for light pulses polarized along  $\vec{b}$ (dashed lines) and $(\vec{a}+\vec{b})$ (solid lines). 
The intensities are color coded as described in figure~\ref{fig2} }
\end{figure}

Furthermore, by shifting weight from $|w_1\rangle$ to $|w_2\rangle$, the excitation transfers charge from
the central atom to the terminal atoms of the trimer.  This charge transfer can be observed by the sudden
drop in the charge correlation in figure~\ref{fig3}.

Via electron-phonon coupling, the charge transfer excites phonons, which are in turn responsible for the
oscillations of the charge and orbital correlation functions in figure~\ref{fig3}. These phonons are closely
related to the long-lived coherent phonons in  $\rm{Pr_{1/2}Ca_{1/2}MnO_3}$\cite{Rajpurohit2020}, which
have been observed experimentally in $\rm{Pr_{1/2}Ca_{1/2}MnO_3}$.\cite{Beaud2014,Esposito2018}
In contrast to the half-doped material,\cite{Sotoudeh2017} these phonons are damped out more rapidly after 
the optically induced phase transition, both for the spin-nematic phase and for the ferroelectric transition.

The photo-induced ferromagnetic coupling between trimers can be understood from the Kanamori-Goodenough
rules \cite{Goodenough1955,Goodenough1958,Kanamori1959} for the majority-spin $|w_2\rangle$ orbital of one trimer
and a majority spin $|w_5\rangle$ Wannier orbital of an adjacent trimer oriented orthogonally. Both orbitals are shown
schematically in figure~\ref{fig1}.f. The two orbitals are linked by an oxygen bridge resulting in an  effective hopping
$t_{eff}=t_0\cos(\gamma)$. This effective hopping depends on the spin angles $\theta$ between both trimers.
The $\gamma$-dependence expresses that the electron hopping is limited to the like-spin components of two spin orbitals. 
The excitation results in a partial occupation of a $|w_2\rangle$ Wannier orbital. Because of the inter-trimer hopping
described above, the final state of the excitation has some weight on the $|w_5\rangle$ Wannier orbital of the adjacent trimer. 
Thus, the excitation causes a charge transfer between the two adjacent trimers, or, in other words, the formation of an inter-trimer bond.
In the ground state, the two Wannier states, $|w_2\rangle$ and $|w_5\rangle$, have a spin angle of $120^\circ$.
Because the electron transfer between  $|w_2\rangle$ and $|w_5\rangle$ Wannier states is limited to the like-spin
component, the spin of the $e_g$ electrons in the $|w_5\rangle$ Wannier state aligns with the excited trimer.
Hund's rule coupling, in turn, leads to a force on the $t_{2g}$ spins $S_R$, which aligns them with the new spin
direction of the $e_g$ electrons. Beyond a critical intensity of the light pulse, provided in table~\ref{tab:critcalintensitues},
the effect is sufficiently strong to drive the system through the phase transition towards the spin-nematic phase of figure~\ref{fig1}.

This description for the optically-induced ferromagnetic coupling is in line with the Goodenough-Kanamori rule
saying that superexchange interactions are antiferromagnetic, if both (spatial) orbitals are half filled, and ferromagnetic,
if they are quarter filled or three-quarter filled.\cite{Goodenough_2008}

\begin{table}[!hbt]
\caption{\label{tab:critcalintensitues}Critical intensities for the optically induced magnetic, charge- and orbital-order phase
 transitions. $A_0$ is the amplitude of the vector potential and $D_p$ is photon absorption in number of absorbed photons per Mn site.}
\begin{center}
    \begin{tabular}{|l|cc|cc|}
    \hline\hline
     polarization &\multicolumn{2}{c|}{$\vec{a}+\vec{b}$, $\vec{a}-\vec{b}$} & \multicolumn{2}{c|}{$\vec{a}$, $\vec{b}$}\\
    unit       & $\hbar/(ea_0)$ & ph/Mn &  $\hbar/(ea_0)$ & ph/Mn \\
    phase transition & $A_0$ & $D_p$ &  $A_0$ & $D_p$ \\
         \hline
    magnetic       & 0.25 & 0.019 & 0.40 & 0.032 \\
    charge order   & 1.00 & 0.088 & 0.65 & 0.097 \\
    orbital order  & 1.60 & 0.168 & 1.20 & 0.158 \\
    \hline\hline
    \end{tabular}
\end{center}
\end{table}

\begin{figure}[t]
\begin{center}
\includegraphics[width=\linewidth]{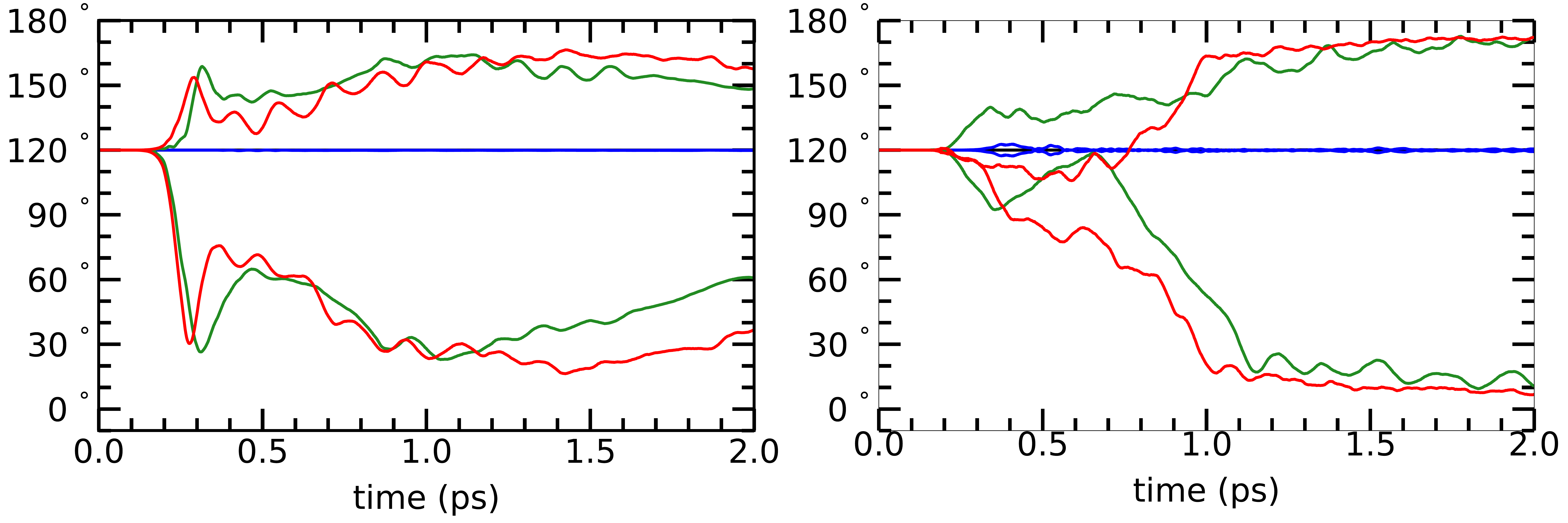}
\end{center}
\caption{\label{fig4}
 Spin angles of adjacent Mn-trimers as function of time for light polarized along $\vec{a}+\vec{b}$ (left) and along $\vec{b}$ (right). 
 The trimers are grouped according to their final spin alignment parallel or antiparallel to the global spin axis.
 On average, the spins deviate by less than $15^\circ$ from the spin axis of the trimer they reside on. The colors
(black, blue, green, red) indicate increasing intensities of the light pulse as quantified in figure~\ref{fig2}.}
  \end{figure}

The co-planar spins of trimers in a chain with spin-angles of $120^\circ$ align ferromagnetically by adding
 an out-of-plane spin component, while reducing the in-plane components as seen from figure~\ref{fig4}. 
The spins reach a ferromagnetic alignment for the first time at approximately 0.3~ps after the light pulse. 
This event is followed by a longer sequence of spin fluctuations on the picosecond time scale. 
The conservation of the total spin results in a complex spin transfer between different chains during this period.

Depending on the polarization, the ferromagnetic chains assume one of two spatial directions, which differ by an angle
of $37^\circ$. A polarization along $\vec{a}+\vec{b}$ leads to a chain with an angle of $18^\circ$ with the $\vec{a}$-axis
and $72^\circ$ with the $\vec{b}$-axis. A polarization along $\vec{a}-\vec{b}$ leads to a phase which is obtained from the
first by an inversion at the Mn$^{3+}$ ion. Interestingly the Mn$^{3+}$ remain at their positions as long as the charge order does not melt.

After having described the dynamics for a polarization along the Mn-trimers, let us now turn to the polarization along $\vec{a}$ or $\vec{b}$.
A light pulse polarized with a $45^\circ$ angle from the trimer axes, results in a quite different behavior.
In this case a ferromagnetic coupling between adjacent orthogonal trimer is established as in the case described above. 
However, each terminal site of a trimer can connect to one of two different orthogonal trimers. 
One might thus anticipate a pattern with chains made of a random sequence of straight sections with 3, 4 and 5 Mn-sites.
However, the simulations only show patterns with segments having 4 Mn-sites. 
This can be attributed to the fact that each such chain already has a dominant spin axis in the ground state, figure~\ref{fig1}.a,
because the spin angles of the trimers within such a chain  alternate between only two rather than three spin orientations with an angle of $120^\circ$. 
The preference of a spin-orientation, build already in the ground state selects a particular pattern, namely four-membered
segments, after the magnetic phase transition.

There are two possible ferroelectric orientations, along $+\vec{a}$ and $-\vec{a}$, which form without apparent preference.
Both patterns break inversion symmetry and are ferroelectric with a polarization vector parallel to the $a$-axis. 
The ferroelectric dipole is formed by the inner two Mn-sites, a formal Mn$^{3+}$ and a formal Mn$^{4+}$ ion, of a four-site segment.

As seen in figure~\ref{fig2} and \ref{fig4}, the spin dynamics leading to the ferroelectric phases is considerably slower than that leading
to the spin-nematic phase. The order parameter for the ferroelectric transition settles after 1~ps as compared to 0.3~ps for the spin-nematic phase.
This may be due to the competition between ferroelectric domains, which have distinct patterns of ferromagnetic chains. Furthermore, the preferred
spin axes of the ferromagnetic chains in the ferroelectric transition form themselfes a spin spiral along the $\vec{b}$-axis. The transition from this
spin spiral to a collinear antiferromangetic order is frustrated. 

The charge and orbital orders are more stable and melt at considerably higher intensities than the optically-induced magnetic transitions
as shown in table~\ref{tab:critcalintensitues}.

The optically-induced magnetic phase transitions discussed here can be investigated experimentally through ultrafast pump-probe
experiments with linearly polarized pulses. The spin dynamics can be accessible via time-resolved resonant soft x-ray
diffraction technique (RXSD) \cite{Ehrke2011}. 

\section{Summary}
In conclusion, our simulations demonstrate the possibility of ultrafast manipulation of the magnetic order by ultra-short (femto-second) light pulses.
The materials studied exhibit a strongly interwined charge-, orbital- and non-collinear spin order. The polarization direction allows
 to drive the system selectively into a spin-nematic phase or a ferroelectric phase.
The particular broken-symmetry state of the spin-nematic phase can be produced selectively by the choice of the polarization.
It can be probed via the resulting anisotropy of the dielectric tensor. The selection of a particular broken-symmetry state of the
ferroelectric phase requires an additional electric field. This field may be static or due to a simultaneous terahertz light pulse with
a phase synchronized with the light pulse. The remanent ferroelectric polarization can be probed via the resulting voltage drop.
Photo-induced magnetic phase transitions predicted here open possibilities for ultrafast manipulation and storage of information
exploiting polarization-sensitive coupling between solid-state and photonic systems. 

\section{Aknowledgement}
The present research work was supported by the Computational
Materials Sciences Program funded by the US Department of Energy,
Office of Science, Basic Energy Sciences, Materials Sciences and
Engineering Division. Financial support from the Deutsche
Forschungsgemeinschaft (DFG, German Research Foundation)
(Grant No 217133147/SFB1073) through Projects B02, B03 and
C03 is gratefully acknowledged. LZT was supported by the Molecular
Foundry, a DOE Office of Science User Facility supported by the
Office of Science of the U.S. Department of Energy under Contract
No. DE-AC02-05CH11231.

\bibliography{ref}
\end{document}